\documentclass[aps,prd,twocolumn,superscriptaddress]{revtex4}
\usepackage{amsfonts,amssymb,amsmath,amscd,graphicx,url,epstopdf,color,verbatim}

\begin{document}

\newcommand{\myvec}[1]{\vec{#1}\,}

\title{Optimal Light Beams and Mirror Shapes for Future
LIGO Interferometers}

\author{Mihai Bondarescu}

\affiliation{Theoretical Astrophysics 130-33, California Institute of Technology, Pasadena, California 91125, USA}
\affiliation{Max-Planck-Institut f\"ur Gravitationsphysik (Albert-Einstein-Institut), Am M\"uhlenberg 1, 14476 Golm, Germany}

\author{Oleg Kogan}
\affiliation{California Institute of Technology, Pasadena, California 91125, USA}

\author{Yanbei Chen}
\affiliation{Theoretical Astrophysics 130-33, California Institute of Technology, Pasadena, California 91125, USA}
\affiliation{Max-Planck-Institut f\"ur Gravitationsphysik (Albert-Einstein-Institut), Am M\"uhlenberg 1, 14476 Golm, Germany}

\date{June 12, 2008}

\begin{abstract}

We report the results of a recent search for the lowest value of
thermal noise that can be achieved in LIGO by changing the shape
of mirrors, while fixing the mirror radius and maintaining a low diffractional loss.  The
result of this minimization is a beam with thermal noise a factor
of 2.32 (in power) lower than previously considered {\em
Mesa Beams} and a factor of 5.45 (in power) lower than the
Gaussian beams employed in the current baseline design. Mirrors
that confine these beams have been found to be roughly conical
in shape, with an average slope approximately equal to the mirror radius divided
by arm length, and with mild corrections varying at the Fresnel
scale.   Such a
mirror system, if built, would impact the sensitivity of LIGO,
increasing the event rate of observing gravitational waves in
the frequency range of maximum sensitivity roughly by a factor of three
compared to an Advanced LIGO using Mesa beams (assuming all other
noises remain unchanged). We discuss  the resulting beam
and mirror properties and study requirements on mirror tilt, displacement
and figure error, in order for this beam to be used in  LIGO detectors. 
\end{abstract}

\maketitle

First-generation laser interferometer gravitational-wave
detectors, such as initial LIGO~\cite{LIGO}, VIRGO~\cite{VIRGO}, GEO\,600~\cite{GEO} and TAMA\,300~\cite{TAMA}, 
have either reached or approached their respective design 
sensitivities, and have taken a first round of coordinated scientific
data. While a detection in the near future is possible with these first data, 
or with upcoming data from Enhanced LIGO, a moderate upgrade of
LIGO detectors, the construction of second generation detectors, 
e.g., Advanced LIGO, which are at least 100 times more sensitive
than first-generation detectors (in power), has already started.  

Advanced LIGO reaches its maximum sensitivity in the frequency range
50-300 Hz.  In this band, the internal thermal noise, i.e., thermal fluctuation
in the mean location of the mirror's surface of reflection, relative to the mirror's  
center of mass, is the dominant noise source.  Depending on location, 
internal thermal noise can be divided
into coating thermal noise and substrate thermal noise; while 
from the physical origin, noise that arise from internal friction 
(viscosity) is called Brownian noise, while noise that corresponds
to thermal damping is called thermoelastic noise.     The dominant 
component of internal thermal noise in fused silica mirrors, the leading 
choice for Advanced LIGO, is the noise contributed by the mirror coating~\cite{GregoryHarry}.

Lowering internal thermal noise will not only directly increase 
LIGO's event rate and thus our chances of seeing gravitational waves
\footnote{for example, a decrease in total noise by a factor of
two corresponds to an increase in event rate of observing
gravitational waves roughly by a factor of three}, but it
may also help bring Advanced LIGO sensitivity beyond the Standard Quantum
Limit. Thus, for the first time, LIGO can study quantum effects as
experienced by 40-kilogram objects \cite{Mueller}.   (Initial LIGO will not benefit 
from lowering internal thermal noise, because in this frequency band it will be limited by 
shot noise and suspension thermal noise.) Other advanced
ground-based detectors as well as Quantum Non-Demolition experiments~\cite{Mueller, Corbitt} will also  benefit
from implementing the research described here when battling mirror
internal thermal noise.

Advanced LIGO's present design (baseline design) uses arm
cavities with Gaussian light beams supported by spherical mirrors ---
with waist-size of Gaussian beams chosen to maintain a $\sim 1\,$ppm
per bounce diffractional loss.  As proposed by Vinet~\cite{Vinet2005}  and Thorne and 
O'Shaughnessy~\cite{OShaughnessyStriginVyatchanin2003},  wider beams supported by non-spherical 
mirrors can average better over thermal fluctuations throughout the 
mirror surface, and therefore give less thermal noise than conventional 
cavities with spherical mirrors and Gaussian beams. 
As a straightforward example, Thorne and O'Shaughnessy proposed the 
so-called Mesa beam, which can lower coating noise by a factor 
of 2.35~\cite{OShaughnessyStriginVyatchanin2003}.

\begin{figure}
\begin{center}
\includegraphics[width=9cm]{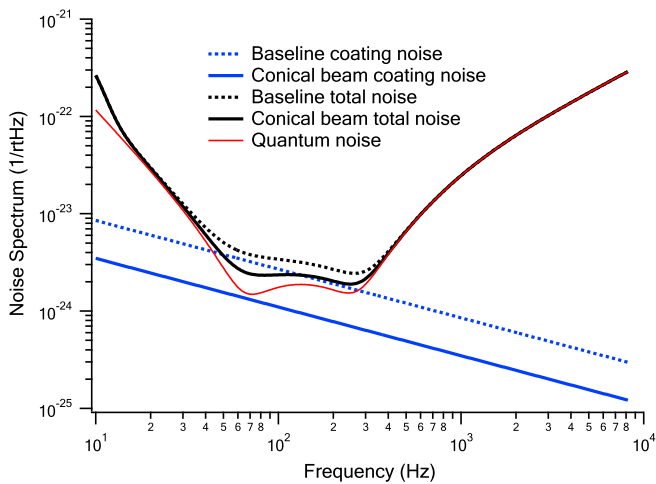}
\end{center}
\caption{Advanced LIGO noise budget for the baseline design
compared a hypothetical situation when the conical mirror is used and
it only impacts coating thermal noise. This situation is not real
because some other noises do change for better or for worse when
one switches to the conical beam.In order to get the true noise
curve of a LIGO-type instrument using conical mirrors, one would
have to do a much more detailed analysis, well beyond the scope of
this paper.} \label{LIGONoise}
\end{figure}

In this work, we aim to search for the beam/mirror configuration with  the 
minimum possible thermal noise.  In order to do so, a systematic 
search over all possible freely propagating beams will be performed.
More precisely, to ensure the completeness of the search, we expand the fundamental
optical mode of the cavity using a Gauss-Laguerre basis (with real-valued coefficients) 
at the center of the cavity, while defining mirror shapes by the beam's phase front at 
the locations of the mirrors. We then use
a gradient-flow minimization method to numerically search for the beam 
profile that corresponds to the minimum thermal noise, subject to the 
constraint of constant diffraction loss. While diffractional loss for any beam/mirror configuration
will be determined from the {\it clipping approximation,}  thermal noise will be
calculated from so-called scaling laws, which state that thermal noise spectral density 
is proportional to a simple spatial-frequency-domain
integral over the Fourier representation of the beam's intensity distribution 
profile on the mirror surface, weighted by a power of the radial wave number.  These
scaling laws apply when the mirrors are approximated as half-infinite (i.e., filling half
of the entire space).  Their existence has been proposed by various authors
(O'Shaughnessy,
Strigin and Vyatchanin \cite{OShaughnessyStriginVyatchanin2003} and Vinet~\cite{Vinet2005} 
for substrate thermoelastic noise,  Vyatchanin \cite{Vyatchanin2004} for
coating Brownian noise, and O'Shaughnessy \cite{OSchaughnessy2006}  for all four
types of noises), while Lovelace \cite{Lovelace} clearly presented and verified the scaling laws for all four
types of noises.

The rest of paper is organized as follows.  Section II
reminds the reader of the Fluctuation-Dissipation-Theorem (FDT) -based approach 
to internal thermal noise due to Levin, and
summarizes its application to the
characterization of spectral density of coating and substrate
noises.  Section~III outlines the minimization problem. 
Section IV describes the results of this minimization - we discuss the mirror profile and
the light beam that correspond to the minimum of substrate and
coating noise.  Section V expands on this discussion and addresses
the issues of tolerance to imperfections and compatibility with
LIGO.  Conclusions are drawn in in Section VI.

While we were working on this paper and after most results were
published in the thesis of MB \cite{DiffFreeBeams}, it came to our
attention that a theoretical lower limit for the coating noise in
an Advanced LIGO detector was derived in \cite{Pierro:2007uh}, 
but it was not proven that it can be reached. 
In this paper we construct an actual beam that approximately achieves this limit.

\section{Noise Characterization}

This section briefly reviews the main results of Levin and
Lovelace, which will be the basis of our work.  Lovelace assumes
half-infinite mirrors, i.e. he neglects all effects arising from a
mirror's finite thickness as well as mirror edge effects, and also
ignores the dynamics of the mirror, e.g., by using the
quasi-static approximation.

LIGO extracts the gravitational wave signal by measuring the position of the mirrors.
The position information is read as $q(t)$, a weighted average of the mirror's
longitudinal position, which depends on  $Z(r, \phi, t)$,  as follows:
\begin{equation}
q(t) \equiv \int_0^{2\pi}d\phi \int_0^{R}dr r p(r)Z(r,\phi, t)
\label{position}
\end{equation}
where $Z$ is the displacement of the mass element at $(r,\phi)$ of
the mirror surface, $R$ is the mirror radius, and $p(r)$ is the
light intensity of the axisymmetric beam at distance $r$ from the
optical axis, which satisfies the normalization of
\begin{equation}
\label{pnorm}
\int_0^R p(r) r dr =1\,.
\end{equation}

Internal thermal noise will cause small fluctuations in the
longitudinal position of the mirror surface, $Z(r, \phi, t)$. This
noise can be divided into two different types: Brownian and
Thermoelastic.  Brownian thermal noise is due to imperfections in
the substrate or coating material that couple normal modes of
vibration to each other. Thermoelastic noise is due to random heat
flow in the mirror that causes some regions to expand and some to
contract. Both noises arise from the substrate as well as from the
mirror coating. Thus, we have to deal with four types of noise:
Coating Brownian noise, Coating Thermoelastic noise, Substrate
Brownian noise and Substrate Thermoelastic noise.

The spectral density, $S$,of the fluctuations in the measured
mirror position, $q$, is derived from the Fluctuation Dissipation
Theorem using Levin's thought experiment \cite{Levin}:
\begin{equation}
S=\frac{2k_BTW_{\rm diss}}{\pi^2fF^2},
\label{Lovelace3}
\end{equation}
where $k_B$ is Boltzmann's constant, $T$ is the mirror
temperature, and $W_{\rm diss}$ is the dissipated power if a
longitudinal force $F$ is applied to the mirror surface with
frequency $f$ and pressure profile $p(r)$, \emph{identical to
light intensity profile}. In initial LIGO, to keep diffraction
losses under 1 ppm per bounce, the beam radius over which $95 \% $
of the signal is collected is kept significantly smaller than the
mirror radius, $R$, and mirror thickness, $H$. Lovelace was forced
to use infinite test-mass approximation because the non-infinite
case is too difficult to solve analytically and will not give a
simple scaling law. He later showed that the infinite test mass
approximation holds reasonably well for beams considerably larger
than in initial LIGO. Fused silica is found to be significantly
less susceptible to the mirror edge effects and finite effects due
to the finite thickness of the mirror than sapphire substrate.  By
using his results, we make the same assumptions in computing the
noise.  As we will see below, the condition that $95\%$ of the
signal is collected from a mirror area smaller than $R$ is very
true in the case of Gaussian beam, almost true in the case of Mesa
beams and almost false in the case of the conical beams proposed
here.

Now, because the resonant frequencies of the mirror are of order
$10^5$ Hz, far higher than the $40 - 200$  range in question , the hypothetical force, $F$, can be
idealized as quasi-static when computing the resulting strain of
the mirror. Advanced LIGO will measure from 10 Hz to 10 kHz;
however, thermal noise is dominant only from 40Hz to 200 Hz.

Thus, to compute the noise $S$, Lovelace \cite{Lovelace}
substitutes in Eq.\ \ref{Lovelace3} the Brownian and Thermoelastic
dissipated power, $W_{\rm diss}$, due to a mirror deformation with the
same pressure distribution as $p(r)$, the light intensity. This
entire procedure is based on Levin's thought experiment
\cite{Levin, Ph237Levin}.  This results in the following
relationships (Eq.~(3.1) of Lovelace \cite{Lovelace}):
\begin{equation}
S_n=A\int_0^\infty dk~k^n| \tilde p(k)|^2,
\end{equation}
where $n=1$ for Coating Brownian and Coating Thermoelastic noise,
$n=0$ for Substrate Brownian noise, $n=2$ for Substrate
Thermoelastic noise, $A$ is a constant that depends on the noise
type and  instrumental setup but does not change with the beam
shape, and $\tilde p(k)$ is the two-dimensional Fourier transform
of the power distribution over the mirror surface, $p(r)$:
\begin{eqnarray}
\tilde p(k) &=&  \int_0^\infty dr~r J_0(kr)p(r) \\
p(r) &=& \int_0^\infty dk~k J_0(kr) \tilde  p(k)
\end{eqnarray}
In the above, $J_0(x)$ is the $0^{th}$ order Bessel function of
the first kind.  This makes numerical evaluation of these noises
easiest in the context of a minimization code that requires noise
to be computed a large number of different power profiles.  In the
case of coating Brownian and Thermoelastic noises, we have $n=1$,
which converts directly to
\begin{equation}
S_1\sim \int_0^\infty dr~r| p(r)|^2
\label{coatingN}
\end{equation}
according to the Parseval Theorem.

Before moving on, we mention that the thermally induced
gravitational wave strain noise power, $S_h(f)$, is related to $S$
by
\begin{equation}
S_h=\frac{4}{L^2}S
\end{equation}
because the interferometer measures
\begin{equation}
h=\frac{(q_1-q_2)-(q_3-q_4)}{L},
\end{equation}
where $q_i$ is the measured position of the $i^{th}$ mirror and
$L=4\,{\rm km}$, the arm cavity length.

\section{The Minimization Problem}
Within the paraxial approximation, Gauss-Laguerre beams 
provide a complete set of orthonormal
basis vectors in the space of all possible LIGO beams propagating
along the $\pm z$ direction, supported by cavities with axi-symmetry
around the $z$ axis.   If we choose to center the cavity at $z=0$, and
place mirrors at $z=\pm L/2$, then for infinite cavities, because of
time-reversal symmetry,  the eigenmodes will have real-valued amplitude 
on the $z=0$ plane, i.e., $U(r,z) \in \mathbf{R}$.  In this way, all 
possible $U$ can be expanded as
\begin{equation}
U(r,z)=\sum_{n=0}^{\infty}A_n\Psi_n(r,z)\,,
\end{equation}
with $A_n \in \mathbf{R}$, and $\Psi_n(r,z)$ the $n$-th Gauss-Laguerre
mode with waist located at $z=0$ (and angular quantum number equal to 0): 
\begin{equation}
\Psi_n(r,z) = \frac{e^{-(2n +
1)i \arctan{\frac{z}{z_0}}}}{w_0 \sqrt{1 +
\left(\frac{z}{z_0}
\right)^2}}\psi_n\left[\frac{\sqrt{2}r/w_0}{\sqrt{1 +
\left(\frac{z}{z_0} \right)^2}}\right]
\end{equation}
where $w_0$ is the waist size of the Gauss Laguerre mode,
$z_0 = \pi w_0^2/\lambda$,   $\psi_n(r) =  Ce^{-r^2/2}\mathcal{L}_n(r^2)$ and
$\mathcal{L}_n$ the  $n$-th Laguerre Polynomial. The Gauss-Laguerre
modes satisfy the following normalization:
\begin{equation}
\int_0^{+\infty}\Psi_j(r,z)\Psi_k(r,z) r dr =\delta_{kj} \,,\quad \forall z\,.
\end{equation}
Note that there is a freedom in choosing Gauss-Laguerre modes with 
different waist sizes, but we choose the Minimal Gauss-Laguerre mode (which has
 the minimum waist size at the mirror location),  
with $z_0 =L/2$.

The infinite mirrors
that support this as the fundamental mode are determined by the
constant-phase surface of $U(r,z)$ around the mirror locations ($z=\pm L/2$).  
The light intensity on the mirror surface is simply
given by \cite{BW}
\begin{equation}
p(r)\propto \left|U(r,z=L/2)\right|^2\,.
\end{equation}
 In the clipping approximation, the mirrors are simply taken as
finite portions of the phase front, with $r \le R$, and the corresponding
diffraction loss is 
\begin{eqnarray}
\epsilon &=& \frac{\int_R^\infty p(r)rdr}{\int_0^\infty p(r)rdr} \nonumber \\
& =& \frac{ \sum_{i,j=0}^{\infty}A_i A_j  \int_R^\infty \Psi_i(r,z_0) \Psi^*_j(r,z_0)  rdr}{\sum_{i=0}^{\infty}A_i^2}
\label{lossconst}
\end{eqnarray}
where $R = 17 {\rm cm}$ is the mirror radius in Advanced LIGO.

Based on this, and the aforementioned scaling law for $S_1$, the
coating thermal noise (\ref{coatingN}) can thus be written as a
quartic function of $A_m$: 
\begin{equation}
S_1 \propto \sum_{i,j,k,l=0}^{\infty}A_i A_j A_k A_l
\int_0^R \Psi_i \Psi^*_j
\Psi_k\Psi^*_l rdr \,,\label{coatingNAm}
\end{equation}
while the normalization of $p(r)$ requires [assuming small 
diffraction loss, Cf.~Eq.~\eqref{pnorm}]
\begin{equation}
 \sum_{i=0}^{\infty}A_i^2=1 
\label{normconst2}
\end{equation}
Fixing an optical loss of $\epsilon$ per bounce adds another constraint,
\begin{equation}
\label{lossconstraint}
\sum_{i,j=0}^{\infty}A_i A_j  \int_R^\infty \Psi_i(r,z_0) \Psi^*_j(r,z_0)  rdr =\epsilon\,.
\end{equation}
Here we choose to have $\epsilon=10^{-6}$.

The minimization is carried out on constraint-satisfying
sub-manifold of the space of linear combinations of Gauss-Laguerre
basis functions. Thus, the gradient generally points out of the
submanifold. To correct for this, at every step of the
minimization after moving along the gradient of the coating noise
(\ref{coatingNAm}), we return to the constraint-satisfying
manifold by moving along the gradient of the diffraction loss
untill (\ref{lossconstraint}) is satisfied. The last step is
renormalizing so that (\ref{normconst2}) is maintained as well.

Numerically, we found that when minimizing (\ref{coatingNAm})
subject to the constraints (\ref{normconst2}) and (\ref{lossconstraint})
in a space with a large number of dimensions, one runs into local
minima. A good way \cite{MihaiEd} to avoid them was to strat in a space 
with few dimensions (low number of Gauss-Laguerre coefficients) and
increase the dimensionality one by one, always using the result of
the previous step as innitial guess.

\section{Results}

\begin{figure}
\begin{center}
\vspace{0.1in}
\includegraphics[width=9cm]{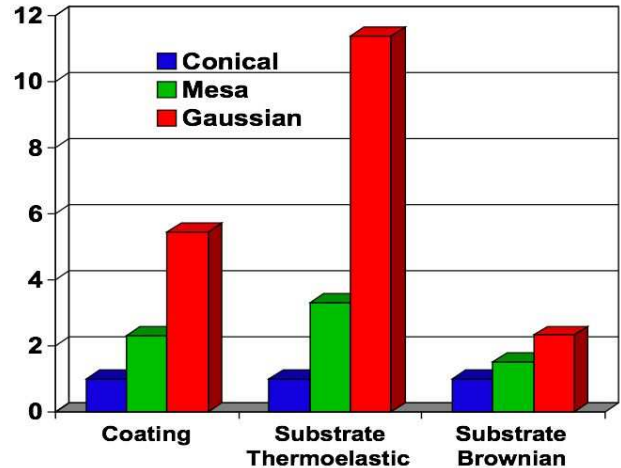}
\vspace{-0.4in}
\end{center}
\caption{The three types of thermal noise for the Gaussian, Conical and Mesa beams normalized so that the conical beam noise is 1 in each cathegory}
\label{NoiseComparison}
\end{figure}

\begin{table}

 \caption{Ratio of Mesa and baseline Gaussian cavity noise to conical cavity noise for different types of noises. 
}
\vspace{0.2in}
 \begin{tabular}{|l|c|c|cl}
 \hline
  & & \\
 Noise & Mesa & Gaussian \\
  & & \\
 \hline
  & & \\
 Coating& 2.32 & 5.45 \\
  & & \\
 \hline
 & & \\
 Substrate Thermoelastic& 3.32& 11.38\\
  & & \\
  \hline
 & & \\
  Substrate Brownian& 1.53& 2.33\\
  & & \\

 \hline
 \end{tabular}

 \label{NoiseDecreaseTable}
 \end{table}

\begin{figure}
\begin{center}
\vspace{-2in}
\includegraphics[angle=90, width=9cm, height=9cm ]{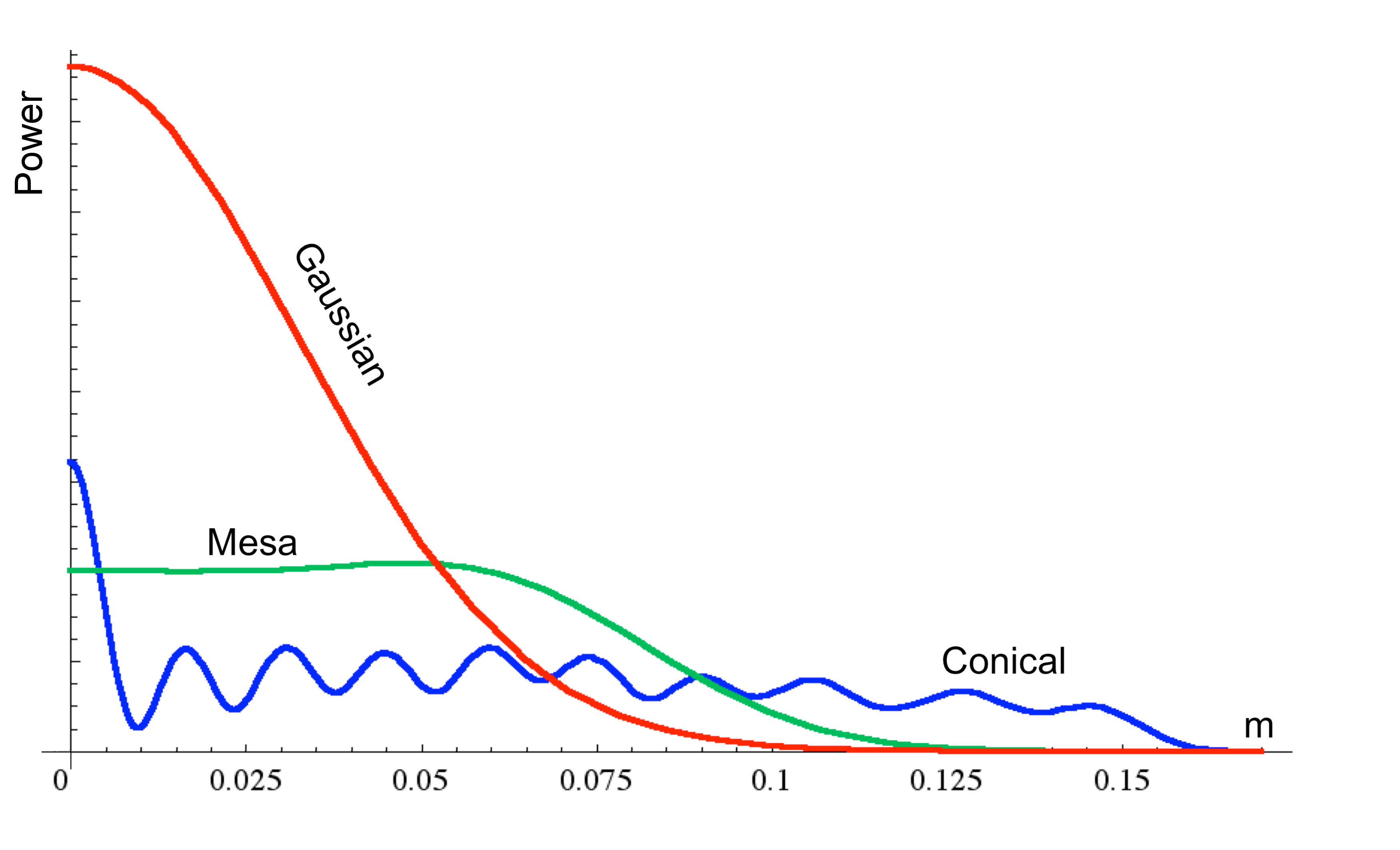}
\end{center}
\vspace{0.5in}
\caption{The power distribution of the lowest-noise 35 coefficient beam found using our minimization algorithm, compared with the previously published Mesa Beam. }
\label{ConePower}
\end{figure}

\begin{figure}
\begin{center}
\vspace{-2in}
\includegraphics[angle=90, width=9cm]{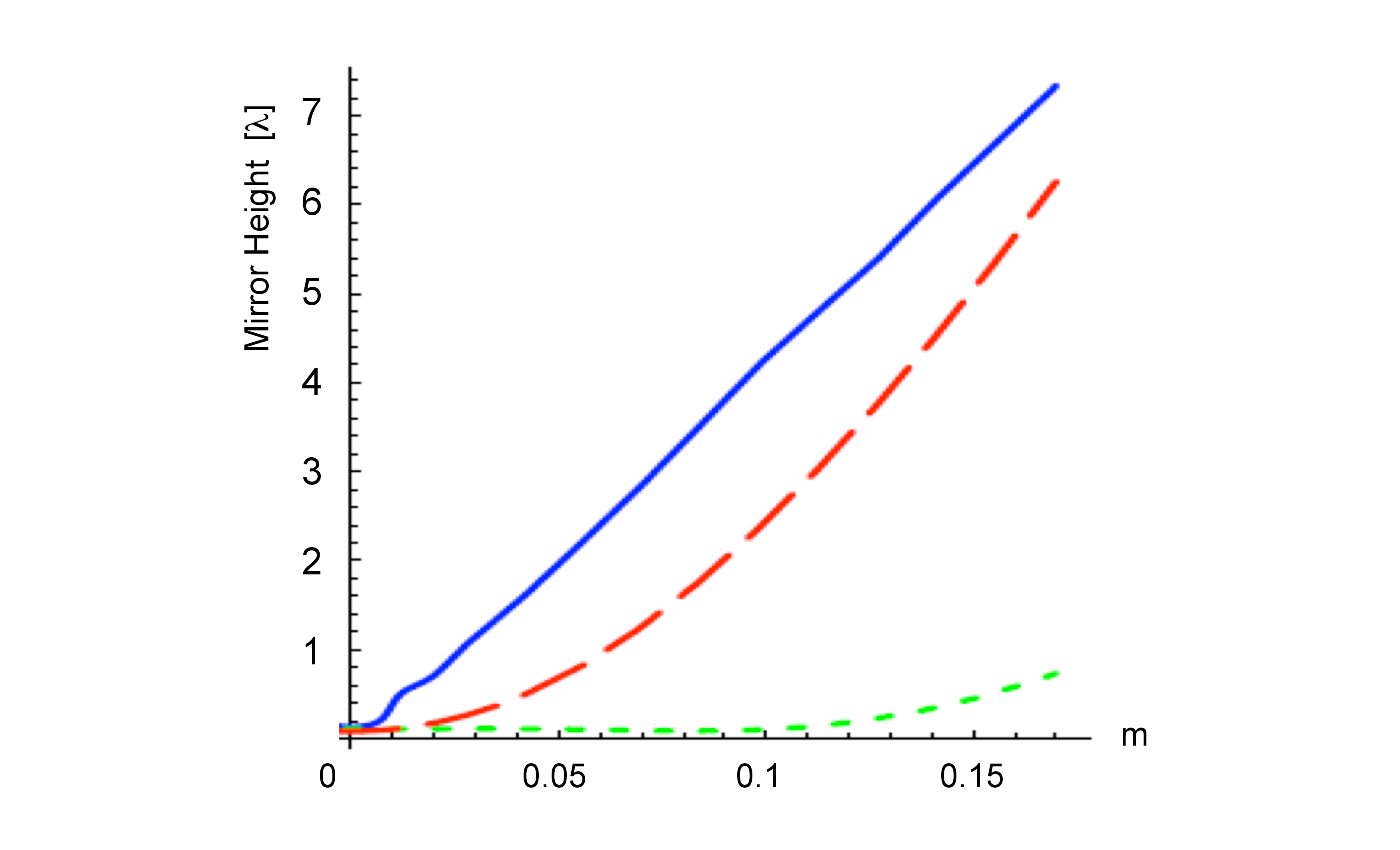}
\vspace{0.5in}
\end{center}
\caption{Mirror height, plotted in units of $ \lambda=1.06 \mu$m, for the nearly conical (solid blue line) mirror compared with the its nearly concentric (dashed red line) and nearly flat (dotted green line) Mesa counterparts. }
\label{ConeComparison}
\end{figure}

The minimization code discussed in the previous section converges
to a beam much wider than Mesa shown in Fig. \ref{ConePower}, while maintaining the same diffraction loss and total power. Switching to a wider beam naturally leads to an overall decrease in all types of thermal noise, even though only the coating noise is actively minimized. This is illustrated in Table \ref{NoiseDecreaseTable} and Fig.  \ref{NoiseComparison}. The mirror profile that should be used to support this beam is shown in Fig.~\ref{ConeComparison}.  Since the mirrors are approximately conical in shape, we will name our beam the {\it Conical beam}.  Interestingly, we note that the mean slope of the cone is roughly $R/L \approx 4.3 \times 10^{-5} \approx 0.4 \lambda/{\rm cm} $.   We also note that the mild deviations from a perfect cone oscillates spatially at the Fresnel scale of $\sqrt{\lambda L/(2\pi)} \approx 2.6\,$cm.

\begin{figure}
\begin{center}
\includegraphics[width=9cm]{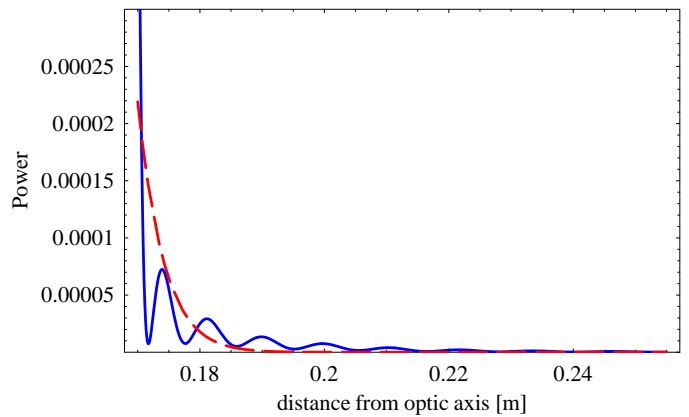}
\end{center}
\caption{The power distribution of the 35 coefficient lowest-noise beam outside the mirror compared to the theoretical prediction for Mesa. In the clipping approximation, the integral of this power is the assumed to be the diffraction loss. }
\label{ConeMesaDiffLoss}
\end{figure}

In Fig.~\ref{ConePower}, when we compare the intensity profile of the Conical beam with those of Mesa and Gaussian beams, we found that the intensity of the Conical beam extends more toward boundaries of the mirrors than Mesa and Gaussian beams -- which increases the level of averaging, and is critical in achieving a significantly lower thermal noise.   This extension does not bring a heightened optical loss, because the Conical beam {\it cuts off} more sharply at the edge of the mirror than Gaussian and Mesa beams, as illustrated in Fig.~\ref{ConeMesaDiffLoss}.  Instead of a Gaussian-like smooth fall-off, power cutoff of the Conical beam is characterized by oscillations outside the mirror radius.  We attribute such an oscillatory cutoff to the mild oscillations of mirror surface shown in Fig.~\ref{ConeComparison}.  Overall, because of extending to larger radii, given the same optical power, the light intensity of a Conical beam everywhere is less then the central area of the baseline Gaussian, while only in the central peak does the Conical beam slightly surpasses the Mesa plateau value.  As a consequence, our beam should be comparable to Mesa in terms of field-strength tolerability by coating materials. 

It is worth mentioning that conical cavities similar to ours have been proposed for the generation of Bessel-Gauss beams~\cite{DurninMiceli}.   However, our beam cuts off more sharply than Bessel-Gauss beams at the edge of the mirrors.  This must also be due to our mirrors' oscillatory deviations from perfect cones.

\subsection{Convergence and Conical Cavities with Fewer Coefficients}

The results presented throughout most of this paper are based on our 35 coefficient minimization result. Low-noise cavities may however be found that employ fewer Gauss-Laguerre coefficients. They are not as good as our final result, but still much better than Mesa. Employing fewer Gauss-Laguerre modes may lead to a cavity that is more desirable for reasons other  than thermal noise. Another reason for showing this figure is to demonstrate convergence.

\begin{figure}
\begin{center}
\includegraphics[width=9cm]{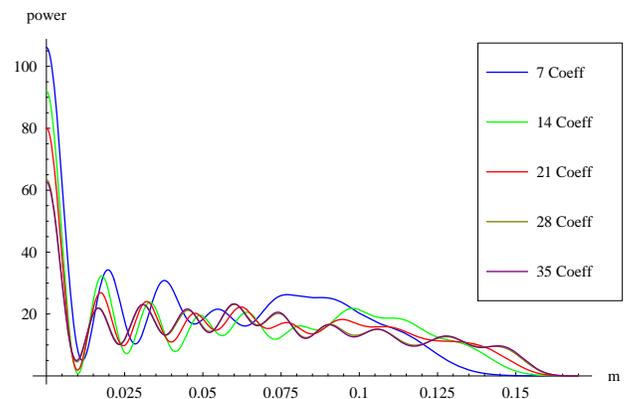}
\end{center}
\caption{Power distribution at the mirror as a function of the number of Gauss-Laguerre coefficients employed in the minimization code. When fewer coefficients are used, the beam is spread over a smaller area of the mirror. }
\label{PowerCoeff}
\end{figure}

\begin{figure}
\begin{center}
\includegraphics[width=9cm]{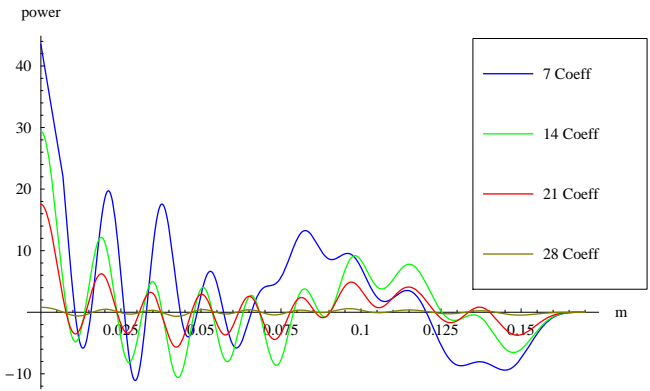}
\end{center}
\caption{The difference in the power distribution at the mirror between the 35 coefficient mode and several other modes with a different number of Gauss-Laguerre coefficients employed in the minimization code.}
\label{PowerCoeffDiff}
\end{figure}

\begin{figure}
\begin{center}
\includegraphics[width=9cm]{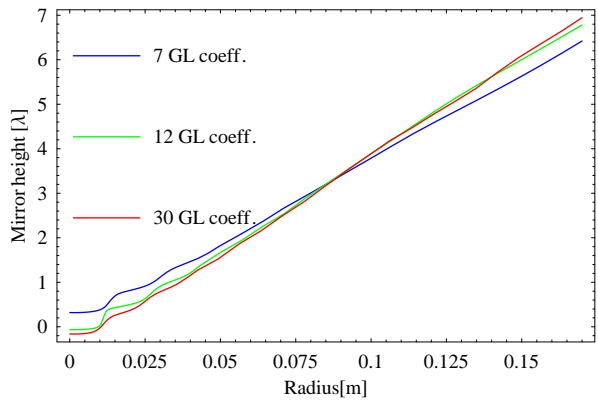}
\end{center}
\caption{Mirrors supporting the 7, 12, and 30 coefficients modes. The mirror height is measured in units of $\lambda$, where $\lambda=1.06 \mu$m is the wavelength of the light used in the interferometer.  }
\label{MirrorCoeff}
\end{figure}

When using fewer Gauss-Laguerre modes in the beam expansion, the beam does not extend all the way out to the end of the mirror. In Fig. \ref{PowerCoeff}, we show the intensity profile at the mirror for beams obtained by minimizing the coating noise over the lowest 7, 14, 21, 27 and 35  Gauss-Laguerre modes. As seen in Fig \ref{MirrorCoeff}, there is little qualitative difference in the mirror shapes, depending on the number of coefficients used. 
\subsection{Optical Modes Supported by Finite Nearly-Conical Mirrors and their Diffraction Loss}

We designed our mirrors to have diffraction losses of 1 ppm in the clipping approximation in order to agree with Mesa and the Baseline Gaussian designs previously considered \cite{BondarescuThorne}.  Here we need to verify that the fundamental mode supported by a finite conical cavity with radius $R$ is indeed close to the beam we constructed, with a diffraction loss close to 1\,ppm.  We are also interested in general the diffraction loss of higher modes.   

\begin{figure}
\begin{center}
\includegraphics[width=9cm]{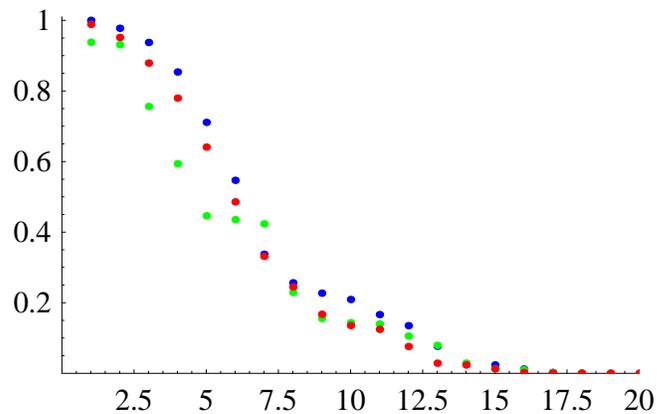}
\end{center}
\caption{Absolute Value of eigenvalues of the one-way propagator. The blue dots represent $n=0$ or axisymmetric modes, the green $n=1$, and the red $n=2$. }
\label{Eigenvalues1}
\end{figure}

Following \cite{ChenSavov}, for modes with angular quantum number $n$, we construct a 1-D radial propagator from one mirror to the other: 
\begin{equation}
K_n(r_1,r_2)=\frac{i^{n+1}k}{L}J_n(\frac{kr_1r_2}{L})e^{ik[h_1(r_1)+h_2(r_2)-L-\frac{r_1^2+r_2^2}{2L}]} ~,
\end{equation}
 where $J_m(z)$ is the $m^{th}$ order Bessed function of the first kind given by
 \begin{equation}
 J_n(z)= \frac{1}{2\pi i^n} \int_0^{2\pi} e^{iz\cos\phi}e^{in\phi}d\phi ~.
 \end{equation}
 $L$ is the length of the arm cavity, $k$ is the wave number $k=\frac{2\pi}{\lambda}$, and $h_{1,2}(r)$ are the mirror heights and are assumed to be equal.   For the conical mirror shape obtained from the previous section, we then computed the eigenvalues of the axisymmetric propagator and obtained only one that was close to 1.  All other higher eigenmodes had high diffraction losses. Figure \ref{Eigenvalues1} we show the absolute values of the  eigenvalues, for modes with $n=0$ (axisymmetric), 1 and 2.  The mode with minimum loss has
 $|\lambda_1|=1-1.45 \cdot 10^{-6}$
 corresponding to a per-bounce diffraction loss of
 \begin{equation}
 1-|\lambda_1^2|=2.9 {\rm ppm}.
 \label{axiloss}
 \end{equation}
 
This is much more than the 1 ppm we find in the clipping approximation and illustrates the limitations of that method.  However, this does not invalidate our configuration because 3\,ppm is still reasonable for use in Advanced LIGO, and when the per-bounce losses of Gaussian and Mesa beams  are relaxed to 3\,ppm, their thermal noises only change by a very small amount.

\section{Tolerance to Imperfections and Compatibility with LIGO}

To investigate the effects of mirror perturbations on the conical cavity, we simulated the 2-D propagation of the light from one mirror of the cavity to the other, using an {\it FFT code} \cite{Erika}, without assuming any symmetry. We used this tool to study the effects of mirror tilt, mirror displacement, and mirror figure errors as well as checking the diffraction losses of the conical cavity.  During light propagation in the cavity, we note that reflection from the mirror is a diagonal operator in position space, while propagation in free space from one plane to the other is diagonal in spatial-frequency domain.  Therefore the most efficient means of propagation is as follows.  When propagating between planes at the mirror locations, use spatial-frequency domain, and when reflecting from mirror surfaces, use position domain --- and insert 2-D FFT between these processes (and hence the name {\it FFT code}).  For a given cavity, we propagate and initial field multiple times in order to obtain the mode with the lowest loss.   As a test, we first used the code on the calculated conical mirror shape, and obtained a beam very close to the theoretical prediction. The diffraction loss per bounce is 3.03 ppm, which agrees well with (\ref{axiloss}). In the following, we use the FFT code in various situations of imperfection, namely mirror tilt, translation and mirror figure error, and estimate the requirement on these imperfections, if the conical cavity is to be used in LIGO. 

 \subsection{Sensitivity to Mirror Tilt}

\begin{figure}
\begin{center}
\includegraphics[width=9cm]{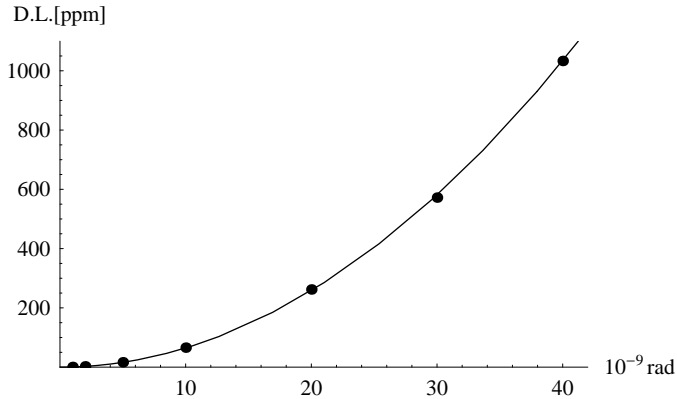}
\end{center}
\caption{ Diffraction loss in a conical cavity as a function of the tilt, when its mirrors are symmetrically tilted.}
\label{TiltInterpolation}
\end{figure}

Our numerical simulation  shows Mesa cavities to increase the diffraction loss to about 3 ppm when perturbed symmetrically by a $10^{-8}$ radians mirror tilt. The same tilt induces a diffraction loss of about 70 ppm in  the conical cavity.   (Antisymmetric tilt can be treated as translation of one mirror, which will be considered next.)  The diffraction loss depends quadratically on the tilt angle as shown in Figure \ref{TiltInterpolation}. The interpolating function, when $x$ is the tilt in radians, is given by
\begin{equation}
{\rm loss}=\left(3.03 + 0.646 \frac{x^2}{\rm{nrad^2}}\right) {\rm ppm}
\end{equation}
As a consequence in order for diffraction loss not to increase significantly due to tile error, the conical mirrors proposed here respond to tilt perturbation more strongly than Mesa does. The suspension and mirror control system would need to be engineered to control the mirror direction better. If LIGO is to use conical mirrors, the tilt needs to be controlled at the level of about 3 nano radians, in order for diffraction loss due to tilt error not to exceed 10\,ppm. 


\subsection{Sensitivity to Mirror Translation}

\begin{figure}
\begin{center}
\includegraphics[width=9cm]{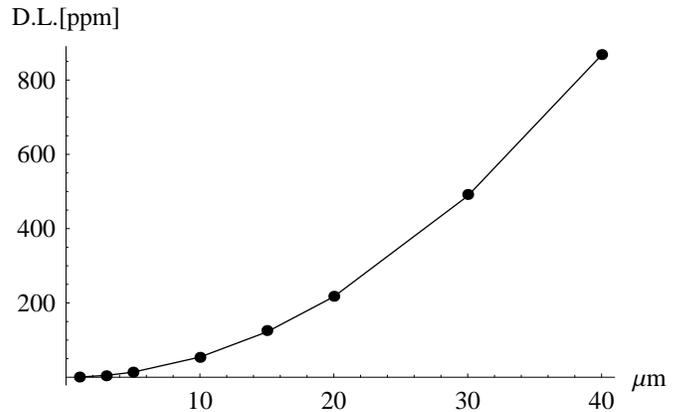}
\end{center}
\caption{Dots represent the diffraction loss in ppm for a conical cavity perturbed by moving one mirror away from the optic axis as a function of this displacement. The continuous line is a quadratic function fit to the data. }
\label{DisplacementLossFit}
\end{figure}.

Simulations show that a conical cavity with a mirror translated by $10~ \mu$m from its ideal position will have a diffraction loss of $57.61$ ppm. A similar diffraction loss is seen in a Mesa cavity with a $4$ mm error in the mirror positioning. Thus the conical cavity is far more sensitive to mirror translation than Mesa. A quadratic function can be fit well  to the as seen in Fig. \ref{DisplacementLossFit}. The diffraction loss in parts per bounce per million of a conical cavity perturbed by moving one of the mirrors a distance $x$ parallel to itself in a direction orthogonal to the optic axis is thus approximatively given by the formula
\begin{equation}
{\rm loss}=\left(3.03 +0.542 \frac{x^2}{\mu {\rm m} ^2}\right) {\rm ppm}
\end{equation}
As a consequence, if Advanced LIGO is to use conical mirrors, mirror position needs to be controlled to about 4 $\mu$m, in order for diffraction loss due to displacement error not to exceed 10\,ppm.  If one uses antisymmetric mirror tilt to compensate errors in mirror location, the tilt would need to be controlled to about $4\,{\mu}\mathrm{m}/4\,{\rm km}=1$ nrad.


 \subsection{Mirror Figure Error and Contribution from Different Scales}
 \label{MirrorFigureError}

 The mirror figure errors are deviations of the mirror surface heights from their theoretical values.  Figure errors have been measured experimentally for the LIGO I mirrors currently used in the experiment.  We used the real measured data from \cite{LIGOFigureError}. Since LIGO I and Advanced LIGO have different mirror sizes, we interpolated the data and stretched it from 12.5\,cm (LIGO-I) to 17\,cm (Advanced LIGO), in the same way as Ref.~\cite{OSchaughnessy2006}.  (This will not be very realistic, because the length scale of the perturbations will make a difference in the losses.)  After numerically solving for the lowest-loss mode using the FFT code, we found that Conical cavities are much more sensitive to this type of figure error, by giving a loss of 405\,ppm, than Mesa cavities which gives a loss of 5\,ppm.  Nevertheless, if we demand a figure error of 1/10 that of the LIGO-I stretched error, we would recover a loss of 5\,ppm, which then becomes reasonable. 
 
 In order to guide further development of mirror manufacturing, it is interesting to study the contribution to optical loss from figure errors at different scales.   Large-scale, or low-spatial-frequency errors cause light to slightly deviate from the cavity axis, while small-scale, or high-spatial-frequency errors will cause light to deviate more significantly from the axis.  In this way, for both Mesa and conical cavities, errors at high spatial frequencies will generate loss anyway, while for Mesa cavities, low-spatial-frequency errors can be less dangerous, since Mesa beam does not cut off so sharply at the edges of the mirrors.  The other way to look at it is that unlike the Conical cavity, the Mesa cavity has more than one low-diffraction-loss modes, which provides a ``reserve'' to maintain a low-loss fundamental mode.  The conical cavity's sensitivity to tilt and displacement agrees with this argument, because tilt and displacement can be considered as low-frequency imperfections.   Following this argument, we also note that since the LIGO-I error is stretched, it tends to decrease the spatial frequency of the figure error, therefore will be biased  {\it against} conical cavities. 

\begin{table}
\begin{tabular}{c|c|c}
\begin{tabular}{c} 
maximum $\lambda$ \\
included
\end{tabular} & Cone & Mesa \\
\hline
$R/16$ & 3.16 & 2.68  \\
$R/8$ & 4.71 & 3.83 \\
$R/4$ & 6.34 & 5.24 \\
$R/2$ & 12.47 & 7.96 \\
$R$ & 56.28 & 5.63 \\
$2R$ & 184.13 & 5.03 \\
$4R$ & 404.83 & 4.86
\end{tabular}
\caption{Optical losses in ppm of conical and mesa cavities, with LIGO-I mirror figure error stretched and high-pass filtered.  When bandwidth of the filter is increased, including more low-frequency (i.e., long length-scale) fluctuations, the loss of the conical cavity increases monotonically, while that of the mesa cavity first increase and then slightly decrease. 
\label{tabfigureerror}}
\end{table}

Now we numerically study contributions from different spatial scales by applying high-pass spatial filters to the stretched LIGO-I noise, and then compute diffraction loss of the resulting fundamental optical mode.  As we see from Table~\ref{tabfigureerror}, as we gradually allow low-spatial-frequency errors to enter, the Conical and Mesa losses initially trace each other --- until a particular spatial scale (in our case $\lambda \sim R$), when Conical loss increases dramatically, eventually climbing up to 100 times that of the Mesa loss, while Mesa loss first increases only mildly, and then even decreases, probably due to particular characteristics of this data set. This means that, in order to make a conical mirror within the LIGO specification, we only need to focus on large-scale errors ($\lambda \stackrel{>}{_\sim} R$) --- while roughness at smaller scales at LIGO-I level is already acceptable for conical cavities. 

\section{Conclusions and further discussions}

We developed and implemented a simple minimization algorithm and used it for finding the LIGO-compatible beam with the lowest thermal noise.  The result of this minimization is a beam with thermal
noise a factor of 2.32 (in power) lower than previously
considered {\em Mesa Beams} and a factor of 5.45 (in power)
lower than the Gaussian beams employed in the current baseline
design.  The mirror that supports this mode is found to have nearly conical shape.  Using an axisymmetric 1-D propagator, we found that contrary to spherical and Mesa cavities, the conical cavity only has one eigenmode with very low diffraction loss (the fundamental mode), while higher modes have much higher optical losses.  

By using an FFT propagation code, we have analyzed the conical cavity's tolerance to practical imperfections. Qualitatively, the conical cavity is much more susceptible than Mesa cavities, to imperfections with low spatial frequencies --- including tilt and mirror translation.  At higher spatial frequencies, the conical cavity is comparable to Mesa cavities.  This behavior can be explained by the fact that the presence of additional low-loss modes allows the Mesa cavity to continue having low diffraction loss when lower-spatial-frequency perturbations are made to the mirrors.  

Recently, Parametric Instability was shown to be a serious problem in Advanced Gravitational Wave Detectors \cite{BSV, Zhao, Miao, Gras}. It arises when the beat frequency of two optical modes is close to the mechanical frequency of an acoustic mode of the mirror. We believe Conical Beams exhibit lower Parametric Instability then Mesa because our higher order modes are very lossy and little power is available to excite the mechanical modes. The topic needs to be thoroughly researched in the future. 

Our numerical simulations show that, in order to achieve diffraction losses close to Mesa on a Conical Cavity, the mirror needs to  be manufactured or corrected with a $CO_2$ Laser such that its large scale deviations from the desired shape are roughly 10 times lower then in the case of Initial LIGO. 
The mirror orientation needs to be controlled to about 1 nanoradian which corresponds to controlling the mirror displacement to the accuracy of several microns. 

Even though the Conical attains the lowest possible coating thermal noise, compromise configurations like those considered in \cite{Lundgren} will be found to provide a reasonable noise level while being easier to implement. 

\begin{acknowledgments}

We thank Kip Thorne for suggesting this problem to us.  We thank 
GariLynn Billingsley for providing information about mirror manufacturing and Rana Adhikari 
for information about suspension systems of future LIGO detectors.  We thank Jayashree Balakrishna, Gregory Daues, Ruxandra Bondarescu, Andrew Lundgren and Dave Tsang for carefully proofreading the manuscript, checking our calculations, and valuable advice on the content and the format of this paper. We thank Kentaro Somiya for helping us with Fig. \ref{LIGONoise}. M.B.~and Y.C.~were supported by the Alexander von Humboldt Foundation's Sofja
Kovalevskaja Programme, NSF grants PHY-0653653 and PHY-0601459, as well as  the David and Barbara Groce startup fund at Caltech. 
MB is grateful for travel support generously offered by 
The Center for Computation and Technology at The Louisiana State University, 
Rochester Institute of Technology, 
The American Physical Society,
The University of Tuebingen,
Universita Degli Studi Del Sanio, 
and Jayashree Balakrishna. 
While preparing this paper, we enjoyed inspiring and stimulating conversations with  
Gabrielle Allen, 
David Blair, 
Manuela Campanelli, 
Erika D'Ambrosio, 
Vincenzo Galdi, 
Kazuaki Kuroda, 
Konstantinos Kokkotas,
Carlos Lousto, 
Vincenzo Pierro,
Innocento Pinto,
Ed Seidel,  Dumitru Vulcanov and   
Hiro Yamanoto. 

\end{acknowledgments}

\end{document}